\documentstyle[preprint,aps]{revtex}
\begin{document}
\preprint{UBCTP-93-022}
\draft
\title{S(k) for Haldane Gap Antiferromagnets: Large-scale Numerical
Results vs. Field Theory and Experiment }

\author{Erik S.\ S\o rensen$^a$ and Ian Affleck$^{a,b}$ }
\address{$^{(a)}$Department of Physics and $^{(b)}$Canadian Institute
for Advanced Research}
\address{
University of British Columbia, Vancouver, BC, V6T 1Z1, Canada}
\date{14-11-93}
\maketitle
\begin{abstract}
The  structure function, $S(k)$, for the $s=1$, Haldane gap
antiferromagnetic chain, is measured accurately using the recent density
matrix renormalization group method, with chain-length 100.  Excellent
agreement with the nonlinear $\sigma$ model prediction is obtained,
both at $k\approx \pi$ where a single magnon process dominates and at
$k\approx 0$ where a two magnon process dominates. We repeat our
calculation with crystal field anisotropy chosen to model NENP,
obtaining
good agreement with both field theory predictions and recent
experiments.  Correlation lengths, gaps and velocities are determined
for
both polarizations.
\end{abstract}
%====================================================================
% BODY OF PAPER

It was first argued by Haldane~\cite{haldane}
that integer-spin $s$, antiferromagnetic
chains exhibit a gap to a triplet magnon above a singlet ground state.
This is by now well confirmed both experimentally~\cite{expgap,renard87}
and numerically~\cite{numgap,tak1,white1,white2,tak93}.
Much insight into these systems can be obtained from the
nonlinear $\sigma$ (NL$\sigma$) model,
the long-distance field theory limit.  Since this
model only becomes an exact representation for large $s$, it is a priori
unclear how accurate it is for $s=1$, the case of greatest experimental
interest. Furthermore, the NL$\sigma$ model itself is not very tractable
and is therefore often approximated by a weakly interacting or free boson
(Landau-Ginsburg) model~\cite{ian}.  The NL$\sigma$ model gives
information about two limiting wave-vectors: $k\to \pi$ and $k\to
0$.  The excitations near $\pi$ are predicted to be a completely stable
single-magnon, plus a three (and higher) magnon continuum.  Near $0$ the
excitations are a two (and higher) magnon continuum.  The
single-magnon contribution to the structure function has a
simple square root Lorentzian (SRL) form near $k\approx \pi$ which only
depends
on the magnon interactions via a renormalization of the overall scale.
On the other hand, near $k\approx 0$ the form of $S(k,\omega )$ depends
quite strongly on the interactions, but the overall scale is
fixed by a sum rule since the total spin operator must obey the spin
commutation relations.  One non-trivial exact result is known about the
NL$\sigma$ model which is of relevance to possible experiments on spin
chains: the explicit form of the two-magnon contribution to
$S(k,\omega)$ at small $k$~\cite{aw}.
The NL$\sigma$ model {\it does not} predict how $S(k,\omega)$
crosses over from one magnon to two magnon behavior as $k$ is swept
across the Brillouin zone.  Anisotropy  may be included in the
NL$\sigma$ model, but this destroys the integrability and we must then
use
the free boson approximation to the two-magnon contribution.

Recently, detailed inelastic neutron scattering experiments~\cite{ma}
were carried out on Ni(C$_2$H$_8$N$_2$)$_2$NO$_2$(ClO$_4$)
(NENP), measuring
$S(k,\omega )$ for $k\geq .3 \pi$.  At all $k$,
the peak width appeared to be resolution limited.
With a ratio of inter- to intra chain coupling estimated~\cite{renard2}
to be $4\times10^{-4}$, NENP is a highly one dimensional compound.
It is, however, not isotropic. The largest of the
anisotropies is the single ion anisotropy, $D$. Neglecting other smaller
anisotropies NENP is well described by the Hamiltonian
\begin{equation}
H=J\sum_i\{{\bf S}_{i}\cdot{\bf S}_{i+1}+D(S_i^z)^2\},
\label{eq:h}
\end{equation}
with $D/J=0.18$ and $J=3.75$meV~\cite{golinelli1}.
Calculations of $S(k,\omega )$ using
exact diagonalization for  $L\leq
18$~\cite{tak93,haas}($L\leq 16$~\cite{golinelli2}) or Monte Carlo for
$L\leq 32$,~\cite{tak93} have found some evidence for the
two-particle nature of the small $k$ excitations from the presence of
spectral weight above the lowest-lying state.

Here, we present accurate results on the equal
time structure function $S(k)$, using the density matrix renormalization
group (DMRG) method recently proposed by White~\cite{white1}, with
chain-length, $L=100$.
Details of our calculation will be given elsewhere~\cite{us2}.
Our results for $S(k)$ are similar to those published
previously for $D/J=0.0$ ($L=18$)~\cite{tak93}, $D/J=0.18$
($L=16$)~\cite{golinelli2} except for a significant
reduction of finite-size effects near $k=\pi$  and the presence of data
at  smaller $k$. This improved precision is crucial for much of our
analysis.
We compare our results on the pure Heisenberg model to the
NL$\sigma$ model predictions.  The SRL form is obeyed
accurately for $k\geq .8 \pi$, with a large renormalization of the
amplitude. The two-magnon NL$\sigma$ model result, including the
overall amplitude, is  remarkably accurate for $k\leq .4 \pi$. Using
previous Monte Carlo results~\cite{tak1} for $\omega_e(k)$,
the minimum energy excitation at
wave-vector $k$, we show that the single mode approximation (SMA) to
$S(k)$ is very good for $k\approx \pi$, indicating that the three (and
higher) magnon contribution is small.  However, the SMA fails  at small
$k$ suggesting the two-particle nature of the excitations there.  We
also obtain results for a single-ion anisotropy term with $D/J=0.18$ to
model NENP.
We determined precise values for the singlet
and doublet gaps,
$\Delta_{||}/J=0.6565(5),$ $\Delta_\perp/J=0.2998(1)$ (in good agreement
with ~\cite{golinelli2}),
correlation lengths,
$\xi_{||}=3.69(5),\ \ \xi_\perp=8.35(7)$
and velocities,
$v_{||}/J=2.38(1),\ \ v_\perp/J=2.53(1)$, and the ground state
energy per spin $e_0/J=-1.2856861$.
These obey the relativistic
relationship, $\xi_a = \hbar v_a/\Delta_a$ to within 2\% for both
polarizations.  Again the SRL form is very accurate
near $k\approx \pi$ and the free boson model is fairly accurate at small
$k$.  The SMA is very accurate at large enough $k$ but fails badly at
small $k$.  Good agreement is  obtained with the experiment apart from
an
overall scale factor of about 1.25 which was to be expected since the
experimental total intensity exceeded the exact sum rule:
$(1/L)\sum_{\alpha
\tilde k}S^{\alpha \alpha}(\tilde k) = s(s+1)$ by 30\% ($\pm$ 30\%).
The behavior of $S(k)$ suggests that
slightly higher resolution and lower $k$ neutron
scattering experiments may reveal the two-particle nature of
the excitations at small $k$.

We implement the DMRG using density matrices of size
$243\times 243$ keeping 81 eigenvectors at each iteration. Our
results are obtained using a finite lattice method on a chain of
length 100. For a discussion of the numerical procedure we refer
the reader to the discussion by White~\cite{white1}. The DMRG yields
a higher accuracy if open boundary conditions are used, and all
of our results are therefore obtained for open ended chains.
If subject to open boundary conditions the $s=1$ Heisenberg
antiferromagnetic
chain has a singlet ground state with positive parity, $0^+$, and
an exponentially low lying triplet, $1^-$~\cite{kennedy,us1}.
In the thermodynamic limit the ground state is thus four fold
degenerate. We have performed our calculations in the
$1^-$ state since this is
computationally more convenient.
We have repeated some of the calculations for the $0^+$ state and
for a smaller system of 60 site. In none of these cases were the results
seen to differ aside from boundary effects which were seen
to diminish with increasing chain length.

Let us first discuss our results for the {\it isotropic} chain
where $D=0.0$ in Eq.~(\ref{eq:h}). The gap
$\Delta/J=0.4107$~\cite{white1,white2,us1} and the velocity
$v/J=2.49(1)$~\cite{us1} is known. Using the DMRG method we calculated
the bulk correlation function $<S^z_{70}S^z_i>$, from which the
correlation
length $\xi=6.03(1)$ can be determined~\cite{us2} in agreement
with previous results~\cite{white2}. Since our calculations are
performed in the $1^-$ state $<S^z_i>$ has a non zero expectation
value giving rise to a disconnected part in the correlation
function. The bulk correlation function we consider is therefore
$<S^z_{70}S^z_i>-<S^z_{70}><S^z_i>$.
We find that end effects are so small
that we may assume this correlation function to be equal to the
correlation function for a periodic chain. The static structure factor,
$S(k)$,
can now be calculated by a simple Fourier transform
$S^{aa}(k)=\sum_r\exp(ikr)S^{aa}(r)$, where $S^{aa}(r)$ is the above
equal time correlation function in real space. Our results for $S=
S^{xx}=S^{yy}=S^{zz}$ are shown in
{}Fig.~\ref{fig:siso} as the open squares. The results obey the sum rule
$(1/L)\sum_{a,k}S^{aa}(k)=s(s+1)$ to within numerical precision up to small
boundary effects.

As already mentioned the isotropic $s=1$ Heisenberg antiferromagnetic
chain
can be described by the NL$\sigma$ model~\cite{haldane}, with
Lagrangian density
\begin{equation}
{\cal L_{\rm nl}}=
\frac{1}{2g}[\frac{1}{v}(\frac{\partial {\bbox{\phi}}}{\partial
t})^2 -v(\frac{\partial {\bbox{\phi}}}{\partial x})^2],
\ \ g=\frac{2}{s},\ \ v=2Jas,
\label{eq:lnl}
\end{equation}
where $a$ is the lattice spacing and ${\bbox{\phi}}$ describes
the sublattice magnetization.
We have approximately ${\bf S_i}=(-1)^is{\bbox{\phi}}+a{\bf l}$,
with
${\bf l}=1/(vg){\bbox{\phi}}\times(\partial {\bbox{\phi}}/\partial t)$.
Since we have set $\hbar=1$, the velocity, $v$, has dimension energy
times length.
The massive triplet of fields ${\bbox{\phi}}$ is restricted to have
unit magnitude, ${\bbox{\phi}}^2=1$. The lowest energy
excitation corresponds
to a single magnon with $k=\pi$ and a relativistic dispersion
$E(k)=\sqrt{(\Delta)^2+(v)^2(k-\pi)^2}$.
Near $k=\pi$ our numerical results for
$S(k)$ have a maximum corresponding to this stable
single magnon, exhibiting the relativistic SRL form:
\begin{equation}
\frac{gv}{2}\frac{1}{\sqrt{\Delta^2+v^2(k-\pi)^2}}.
\label{eq:srl}
\end{equation}
Excellent agreement between this form and the numerical results is seen
for $k/\pi\geq 0.8$ (see Fig.~\ref{fig:siso}).
Since $v$ and $\Delta$ are known from independent
calculations this allows us to determine the coupling constant, $g\sim 1.28$.
This result can be compared to the usual large $s$ result of $g=2/s=2$ and
tentatively to the $1/s$ expansion result~\cite{chubokov}
$g\sim 1.44$ for the bare coupling.

Near $k=0$ $S(k)$ approaches zero as expected since the ground state is
a singlet. In this region $S(k)$ is expected to be dominated by  a two (and
higher) magnon continuum~\cite{tak1} corresponding to the excitation
of two magnons with momenta $\pm\pi$.
Exact results for the two magnon
contribution to $S(k,\omega)$ within the NL$\sigma$ model frame work
have been obtained in Ref.~\onlinecite{aw}:
\begin{equation}
a^2|G(\theta)|^2\frac{vk^2}{2\pi}\frac{\sqrt{\omega^2-(vk)^2
-4\Delta^2}}{(\omega^2-(vk)^2)^{3/2}},\ \ \omega^2-(vk)^2>4\Delta^2.
\label{eq:sknl}
\end{equation}
Here $\omega^2-(vk)^2=4\Delta^2{\rm cosh}^2(\theta/2)$ and
$|G(\theta)|^2$
is given by the expression
\begin{equation}
|G(\theta)|^2=\frac{\pi^4}{64}\frac{1+(\theta/\pi)^2}{1+(\theta/2\pi)^2}
\left(\frac{{\rm tanh}(\theta/2)}{\theta/2}\right)^2.
\end{equation}
This result can be numerically integrated over $\omega$ to yield
the two magnon contribution to the static structure factor. The result
is shown as the solid line in Fig.~\ref{fig:siso} and, as can be
seen, it is remarkably accurate out to $k/\pi\leq 0.4$.

If the
nonlinear constraint, ${\bbox{\phi}}^2=1$, on the field $\bbox{\phi}$
in the NL$\sigma$ model is lifted, a simpler model of
almost free bosons can be arrived at~\cite{ian} with Lagrangian density
\begin{equation}
{\cal L}=\sum_{i=1}^3\Big[\frac{1}{2v_i}(\frac{\partial\phi_i}{\partial
t})^2
-\frac{v_i}{2}(\frac{\partial\phi_i}{\partial x})^2
-\frac{\Delta_i^2}{2v_i}(\phi_i)^2\Big]-\lambda{\bbox{\phi}}^4.
\end{equation}
By allowing the 3 velocities and gaps to be phenomenologically
adjusted, anisotropy can be included.
If the $\lambda{\bbox{\phi}}^4$ term is neglected a mean field
theory of free bosons is obtained.
In the isotropic case,
within this free boson model, the two magnon contribution to $S(k)$ can
be shown~\cite{aw} to be equal to Eq.~(\ref{eq:sknl}) with $|G(\theta)|^2\equiv
1$.
The free boson estimate is shown as the long dashed line in
{}Fig.~\ref{fig:siso}. While in good qualitative agreement with the
numerical results near $k=0$ the discrepancy at larger $k$, between
this result and the exact NL$\sigma$ prediction,
demonstrate the importance of interaction effects that are neglected
in the free boson theory.

Within
the SMA the dynamical structure factor $S(k,\omega)$ is
approximated by a $\delta$-function at the single
mode frequency
$\omega_{\rm SMA}$,
$S(k,\omega)=S_0(k)\delta[\omega-\omega_{\rm SMA}(k)]$.
Since the first moment of the dynamical structure factor
obeys a sum rule~\cite{hohenberg} this implies~\cite{us2} that
\begin{equation}
\omega_{\rm SMA}^{zz}(k)=-J(F_x+F_y)(1-\cos(k))/S^{zz}(k),
\end{equation}
with $F_a=<S^a_iS^a_{i+1}>$.
Denoting the lower edge of the spectrum by $\omega_e$,
one can show~\cite{tak88,us2} that the inequality
$\omega_{\rm SMA}(k)\geq\omega_e(k)$ is obeyed.
{}Furthermore it can be shown~\cite{us2}
that if
$S(k,\omega)$ has a $\delta$-function peak at the frequency
$\omega_\delta(k)$, which is {\it below} the lower edge
of the {\it continuos} part of the spectrum, $\omega_c(k)$, then
$\omega_{\rm SMA}(k)\geq\omega_\delta(k)$.
This inequality breaks
down if $\omega_\delta(k)>\omega_c(k)$. Using the numerical results
$F_a=e_0/3=-0.4671613$ for the isotropic chain we can now
calculate $\omega_{\rm SMA}(k)$ from $S(k)$. Our results are shown in
{}Fig.~\ref{fig:omiso} along with the quantum Monte Carlo (QMC) results
of Takahashi~\cite{tak1} for the lower edge, $\omega_e$, of the
spectrum. Also shown is the relativistic dispersion relation
(solid line),
which should be very accurate near $k=\pi$, and the bottom
of the two-magnon continuum (long dashed line),
$2\sqrt{\Delta^2+v^2(k/2)^2}$, valid
near $k=0$. As can
be seen in Fig.~\ref{fig:omiso} both predictions coincides with the
QMC results.
While highly accurate near $k=\pi$ the SMA is clearly
seen to break down near $k=0$.
{}From our results for the structure factor we
expect the two magnon scattering to become important around
$k\sim 0.4\pi$ which seems to be consistent with the divergence of
$\omega_{\rm SMA}$ from the QMC results. If compared to the relativistic
dispersion, which should represent the single magnon dispersion
relation $\omega_k$ near $k=\pi$ very accurately, it is seen that
also here there must be a small
multi-magnon contribution since $\omega_{\rm SMA}$
is {\it above} $\omega_k$.

We now turn to a discussion of our results for the anisotropic chain
with $D/J=0.18$.
Using the DMRG we have calculated $<S^z_{50}S^z_i>$ and
$<S^x_{50}S^x_i>$ in the $1^-$ state. In this case the disconnected part,
$<S^a_{50}><S^a_i>$,
is so small that we can neglect it. Fourier transforming we obtain
$S^{||}=S^{zz}$ and $S^{\perp}=S^{xx}=S^{yy}$, respectively. Our
results are
shown in Fig.~\ref{fig:sanis}. We estimate end effects to give an error
of 0.05\% in $S^{\perp}(k=\pi)$, somewhat larger at $k=0$. End effects
are judged to be negligible for $S^{||}$. Also shown in
{}Fig.~\ref{fig:sanis} are the INS results of Ma et al.~\cite{ma}
which are directly comparable to our results.
The open triangles are points where to within experimental accuracy
$S^{||}$ and $S^{\perp}$ were identical. The full squares and
circles are data points where $S^{||}$ and $S^\perp$, respectively,
could be resolved experimentally.
At $k=\pi$ the experimental results are about 20-30\% higher than
our numerical results. Taking into account a 25\%~\cite{reich}
uncertainty in the overall (multiplicative) normalization of the
experimental results we find a good agreement between the
numerical and experimental results.

Allowing for different velocities,
gaps and coupling constants
for the two modes we expect $S^{aa}(k)$ to be
well described by the SRL form Eq.~(\ref{eq:srl}) near $k=\pi$.
With the gaps and velocities mentioned above and with
$g_{||}=1.44$, $g_{\perp}=1.17$ this SRL form is shown
as the dashed lines in Fig.~\ref{fig:sanis}. Very good agreement
when $k\geq 0.8\pi$ between the SRL form and the numerical results
is evident for both $S^{||}$ and $S^{\perp}$.

In the region near $k=0$ we see that since rotational symmetry
around the x and y axis is broken, $S^\perp$ can take on a non zero value
at $k=0$. This is clearly seen in the numerical results for
$S^{\perp}$.
We can extend the results of Ref.~\onlinecite{aw} for the free
boson prediction to allow for different velocities.
{}For the two magnon contribution to $S^{aa}(k)$ near $k=0$ we
obtain~\cite{us2}
\begin{equation}
a^2\int \frac{dk'dk''}{8\pi}(\frac{\omega^b_{k'}v_c}{\omega^c_{k''}v_b}+
\frac{\omega^c_{k''}v_b}{\omega^b_{k'}v_c}-2)\delta(k-k'-k''),
\label{eq:spar}
\end{equation}
with cyclic permutation of the indices $a,b,c$.
Here $\omega^a_k=\sqrt{\Delta_a^2+(v_ak)^2}$.
These results are shown as the solid lines in Fig.~\ref{fig:sanis}.
As was the case for the isotropic chain the free boson prediction seems
to be qualitatively correct near $k=0$.
The crossings of $S^{||}(k)$ and $S^\perp(k)$ around
$k=0.85\pi$ and $0.1\pi$ (see Fig..~\ref{fig:sanis}) are
reproduced correctly by the free boson model.
In the range $0.1\pi<k<0.85\pi$
$S^{||}$ is the
larger of the two structure factors.

We have repeated the SMA calculation for the anisotropic
case finding qualitatively similar behavior as for the isotropic chain.
Again the SMA fails for small $k$ while giving relatively
precise results near $k=\pi$, indicating a small
but non zero contribution from the multi magnon continuum at $k=\pi$.

We would like
to thank Steven White for sendings us Refs. \onlinecite{white1,white2}
prior to publication as well as for useful discussions.
We are also indebted to
S.~Ma, C.~Broholm, D.~H.~Reich, B.~J.~Sternlieb, and R.~W.~Erwin
for permission to present their data. ESS thanks Dan Reich
for several helpful discussion and for suggesting looking at the
single mode approximation for this problem.
This research was supported in part
by NSERC of Canada.

\begin{figure}
\caption{The structure factor, $S$, vs. $k/\pi$ for
a 100 site {\it isotropic} chain. The numerical results are shown
as open squares.
}
\label{fig:siso}
\end{figure}

\begin{figure}
\caption{Dispersion of excitations
vs. $k/\pi$ for
a 100 site {\it isotropic} chain.
The points are QMC data from Ref.~\protect\onlinecite{tak1} for $L=32$.
The short dashed line is $\omega_{\rm SMA}(k)$ obtained from $S(k)$ for
$L=100$.
}
\label{fig:omiso}
\end{figure}

\begin{figure}
\caption{$S^{||}$ and $S^{\perp}$
vs. $k/\pi$ for
a 100 site {\it anisotropic} chain. The numerical results are shown
as open squares and circles, respectively.
The data points are INS data from Ref.~\protect\onlinecite{ma}.
}
\label{fig:sanis}
\end{figure}


\begin{references}

\bibitem{haldane}
{}F.~D.~M.~Haldane, Phys.~Lett.\ {\bf 93A}, 464 (1983);

\bibitem{expgap}
W.~J.~L.~Buyers, R.~M.~Morra, R.~L.~Armstrong, M.~J.~Hogan,
P.~Gerlach, and K.~Hirikawa, Phys.\ Rev.\ Lett. {\bf 56},
371 (1986).

\bibitem{renard87}
J. P. Renard, M. Verdaguer, L. P. Regnault, W. A. C. Erkelens,
J. Rossat-Mignod, and W. G. Stirling, Europhys.\ Lett.\ {\bf 3},
945 (1987).

\bibitem{numgap}
R.~Botet and R.~Julien, Phys.\ Rev. B {\bf 27}, 613, (1983);
R.~Botet, R.~Julien, and M.~Kolb, Phys.\ Rev.\ B {\bf 28}, 3914 (1983);
M.~Kolb, R.~Botet, and R.~Julien, J.\ Phys.\ A {\bf 16} L673 (1983);
M.~P.~Nightingale and H.~W.~J.~Bl\" ote, Phys.\ Rev.\ B {\bf 33},
659 (1986); H.~J.~Schulz and T.~A.~L.~Ziman, Phys.\ Rev.\ B {\bf 33},
6545 (1986).

\bibitem{tak1}
M.~Takahashi, Phys.\ Rev.\ Lett.\ {\bf 62}, 2313 (1989).

\bibitem{white1}
S.~R.~White, Phys.\ Rev.\ Lett. {\bf 69}, 2863 (1992).

\bibitem{white2}
S.~R.~White and D.~A.~Huse,
Phys.\ Rev.\ B {\bf 48}, 3844 (1993).

\bibitem{tak93}
M. Takahashi, Phys.\ Rev.\ B {\bf 48}, 311 (1993).

\bibitem{ian}
%prl on ginzburg-landau
I.~Affleck, Phys.\ Rev.\ Lett. {\bf 62}, 474, E 1927 (1989);
{\bf 65}, 2477, 2835 (1990);
Phys.\ Rev.\ B {\bf 41}, 6697 (1990);
{\bf 43}, 3215 (1991).

\bibitem{aw}
I.~Affleck and R.~A.~Weston, Phys.\ Rev.\ B {\bf 45}, 4667 (1992).

\bibitem{ma}
S.~Ma, C.~Broholm, D.~H.~Reich, B.~J.~Sternlieb, and R.~W.~Erwin,
Phys.\ Rev.\ Lett.\ {\bf 69}, 3571 (1992).

\bibitem{renard2}
J. P. Renard, M. Verdaguer, L. P. Regnault, W. A. C. Erkelens,
J. Rossat-Mignod, J. Ribas, W. G. Stirling, and C. Vettier,
J.\ Appl.\ Phys.\ {\bf 63}, 3538 (1988); J. P. Renard, V. Gadet,
L. P. Regnault, and M. Verdaguer, J. Mag. Mag. Mat. {\bf 90-91},
213 (1990).

\bibitem{golinelli1}
O.~Golinelli, Th.~Jolic\oe ur, and R.~Lacaze,
Phys.\ Rev.\ B {\bf 45}, 9798 (1992).

\bibitem{haas} S. Haas, J. Riera and E. Dagotto, Phys.\ Rev. B {\bf
48}, 3281 (1993).

\bibitem{golinelli2}
O.~Golinelli, Th.~Jolic\oe ur, and R.~Lacaze,, J. Phys. \ Cond.\ Matt.
{\bf 5}, 1399 (1993).

\bibitem{us2}
E.~S.~S\o rensen and I.~Affleck,
unpublished.

\bibitem{kennedy}
T.~Kennedy, J.~Phys.\ Cond.\ Matt.\ {\bf 2}, 5737 (1990).

\bibitem{us1}
E.~S.~S\o rensen and I.~Affleck,
Phys.\ Rev.\ Lett. {\bf 71}, 1633 (1993).

\bibitem{chubokov}
D.~V.~Kveshchenko and A.~V.~Chubukov,
Zh.\ Eksp.\ Teor.\ Fiz.\ {\bf 93}, 1904 (1987)
[Sov.\ Phys.\ JETP {\bf 66}, 1088 (1987)].

\bibitem{hohenberg}
P.~C.~Hohenberg and W.~F.~Brinkman, Phys.\ Rev.\ B {\bf 10},
128 (1974).

\bibitem{tak88}
M.~Takahashi, Phys.\ Rev.\ B {\bf 38}, 5188 (1988).

\bibitem{reich}
D.~H.~Reich, private communication.

\end{references}
\end{document}